\documentclass[aip,rsi,reprint,graphicx]{revtex4-1}
\usepackage{amssymb}
\usepackage[latin1]{inputenc}
\usepackage{amsmath}
\usepackage{graphicx}
\usepackage{color}
\usepackage[english]{babel}
\usepackage{dcolumn}
\usepackage{bm}

\begin{document}
\frenchspacing

\title{Microwave soft x-ray microscopy for nanoscale magnetization dynamics in the 5--10 GHz frequency range}

\author{Stefano Bonetti}
\email{bonetti@slac.stanford.edu}
\altaffiliation{Current affiliation: Dept. of Physics, Stockholm University, Stockholm, Sweden.}
\affiliation{\mbox{Department of Physics, Stanford University, Stanford, CA, USA}}
\affiliation{\mbox{Stanford Institute for Materials and Energy Science, SLAC National Accelerator Laboratory, Menlo Park, CA}}

\author{Roopali Kukreja}
\affiliation{\mbox{Department of Materials Science and Engineering, Stanford University, Stanford, CA, USA}}
\affiliation{\mbox{Stanford Institute for Materials and Energy Science, SLAC National Accelerator Laboratory, Menlo Park, CA}}

\author{Zhao Chen}
\affiliation{\mbox{Department of Physics, Stanford University, Stanford, CA, USA}}
\affiliation{\mbox{Stanford Institute for Materials and Energy Science, SLAC National Accelerator Laboratory, Menlo Park, CA}}

\author{Detlef Spoddig}
\affiliation{\mbox{Institut f{\"u}r Experimentalphysik, Universit{\"a}t Duisburg-Essen, Germany}}
\author{Katharina Ollefs}
\affiliation{\mbox{Institut f{\"u}r Experimentalphysik, Universit{\"a}t Duisburg-Essen, Germany}}
\affiliation{\mbox{European Synchrotron Radiation Facility, 38043 Grenoble Cedex, France}}
\author{Christian Sch{\"o}ppner}
\affiliation{\mbox{Institut f{\"u}r Experimentalphysik, Universit{\"a}t Duisburg-Essen, Germany}}
\author{Ralf Meckenstock}
\affiliation{\mbox{Institut f{\"u}r Experimentalphysik, Universit{\"a}t Duisburg-Essen, Germany}}
\author{Andreas Ney}
\affiliation{\mbox{Institut f{\"u}r Experimentalphysik, Universit{\"a}t Duisburg-Essen, Germany}}
\affiliation{\mbox{Solid State Physics Division, Johannes Kepler University, 4040 Linz, Austria}}

\author{Jude Pinto}
\affiliation{\mbox{Linear Coherent Light Source, SLAC National Accelerator Laboratory, Menlo Park, CA}}
\author{Richard Houanche}
\affiliation{\mbox{Linear Coherent Light Source, SLAC National Accelerator Laboratory, Menlo Park, CA}}
\author{Josef Frisch}
\affiliation{\mbox{Linear Coherent Light Source, SLAC National Accelerator Laboratory, Menlo Park, CA}}

\author{Joachim St{\"o}hr}
\affiliation{\mbox{Stanford Institute for Materials and Energy Science, SLAC National Accelerator Laboratory, Menlo Park, CA}}
\author{Hermann D{\"u}rr}
\affiliation{\mbox{Stanford Institute for Materials and Energy Science, SLAC National Accelerator Laboratory, Menlo Park, CA}}

\author{Hendrik Ohldag}
\affiliation{\mbox{Stanford Synchrotron Radiation Laboratory, SLAC National Accelerator Laboratory, Menlo Park, CA}}

\begin{abstract}
We present a scanning transmission x-ray microscopy setup combined with a novel microwave synchronization scheme in order to study high frequency magnetization dynamics at synchrotron light sources. The sensitivity necessary to detect small changes of the magnetization on short time scales and nanometer spatial dimensions is achieved by combination of the developed excitation mechanism with a single photon counting electronics that is locked to the synchrotron operation frequency. The required mechanical stability is achieved by a compact design of the microscope. Our instrument is capable of creating direct images of dynamical phenomena in the 5-10 GHz range, with 35 nm resolution. When used together with circularly polarized x-rays, the above capabilities can be combined to study magnetic phenomena at microwave frequencies, such as ferromagnetic resonance (FMR) and spin waves. We demonstrate the capabilities of our technique by presenting phase resolved images of a $\sim6$ GHz nanoscale spin wave generated by a spin torque oscillator, as well as the uniform ferromagnetic precession with $\sim0.1$ deg amplitude at $\sim9$ GHz in a micrometer-sized cobalt strip.
\end{abstract}

\maketitle

\section{Introduction}

Modern science requires instrumentation that allows for investigating ever smaller spatial dimensions and increasingly faster time-scales. In magnetism research, this requirement has been pushed in the last decade by the prediction and the realization of current induced magnetization dynamics, achieved either via spin transfer torque or spin-orbit effects, such as the spin-Hall effect. However, while it is nowadays possible to map magnetism at the nanoscale and to measure magnetization dynamics at fast time scales, it is still challenging to combine high spatial and temporal resolution in the same measurement.

In the last decade, a few groups around the world started to bridge this gap using x-rays generated at synchrotron light sources. Such x-rays are in principle capable of looking at magnetism both at the nanoscale, given their short wavelength, and at fast time scales, since a typical synchrotron x-ray pulse has a duration of 50-100 ps. Some work has been done on measuring ferromagnetic resonance (FMR) in extended samples \cite{Arena:ReviewOfScientificInstruments:2009,Boero:RevSciInstrum:2009,Buschhorn:JSynchrotronRadiat:2011,Cheng:JournalOfAppliedPhysics:2012,Goulon:IntJMolSci:2011,Kaznatcheev:AipConferenceProceedings:2011,
Rogalev:JournalOfInfraredMillimeterAndTerahertzWaves:2012,Warnicke:JournalOfAppliedPhysics:2013,bailey2013detection,stenning2015magnetization}, but measurements combining the time-resolved probing of fast magnetization dynamics with nanoscale x-ray microscopy are challenging and still very rare\cite{PhysRevLett.106.167202,APL.101.182407,kammerer2011magnetic,bisig2013correlation,kim2014synchronous}. Such studies have also been restricted to dynamics at frequencies lower than 3 GHz. This is an important constraint, which greatly limits the type of samples that one can look at. Indeed, many of the materials and of the phenomena of interest in modern magnetism show dynamics in the 5--10 GHz range. 

In this work, we describe a scanning transmission soft x-ray microscope that has been optimized for the study of magnetization dynamics both in the 5--10 GHz range and with 35 nm resolution. In order to achieve this, we designed microwave synchronization electronics that can be synchronized with RF frequency of the Stanford Synchrotron Radiation Lightsource (SSRL). 
Our measurement technique relies on a quasi-stroboscopic detection scheme, which allows us to compensate for typical drifts associated with x-ray absorption measurements at a synchrotron. In addition, the mechanical components of the microscope and the x-ray illumination optics have been optimized for long term stability. This allows for acquiring images with long integration times, in turn enabling the detection of extremly small transient magnetization changes. We demonstrate the capabilities of our instrument by recording the magnetization dynamics present in the spin waves emitted by a nanocontact spin torque oscillator at around 6.3 GHz, and the ferromagnetic resonance of $\sim0.1$ deg amplitude in a micron-sized lithographic element at 9.1 GHz.


\section{Experimental setup}

The measurement setup comprises of three parts. The first part is the scanning transmission x-ray microscopy (STXM) instrument at beamline 13-1 at the Stanford Synchrotron Radiation Lightsource (SSRL). The second part is a single-photon counting detector synchronized with the RF frequency of SSRL. This detector has been developed in the past in our group\cite{RSI.78.014702}, and in the following subsection we briefly describe its functionality and the modification that we introduced. The third part, discussed in the last subsection, is a microwave synchronization board that is able to produce a single frequency microwave signal phase-locked to the synchrotron frequency. This microwave signal is then fed into the sample to drive a dynamics that is in turn phase-locked to the synchrotron and to the detector. Thus, our instrument resembles a setup for ``pump-probe'' type of experiments, where the microwave signal act as the ``pump'', and the x-rays as the ``probe''.

\subsection{X-ray Source and Microscope}

\begin{figure}[b]
\centering
\includegraphics[width=0.5\textwidth]{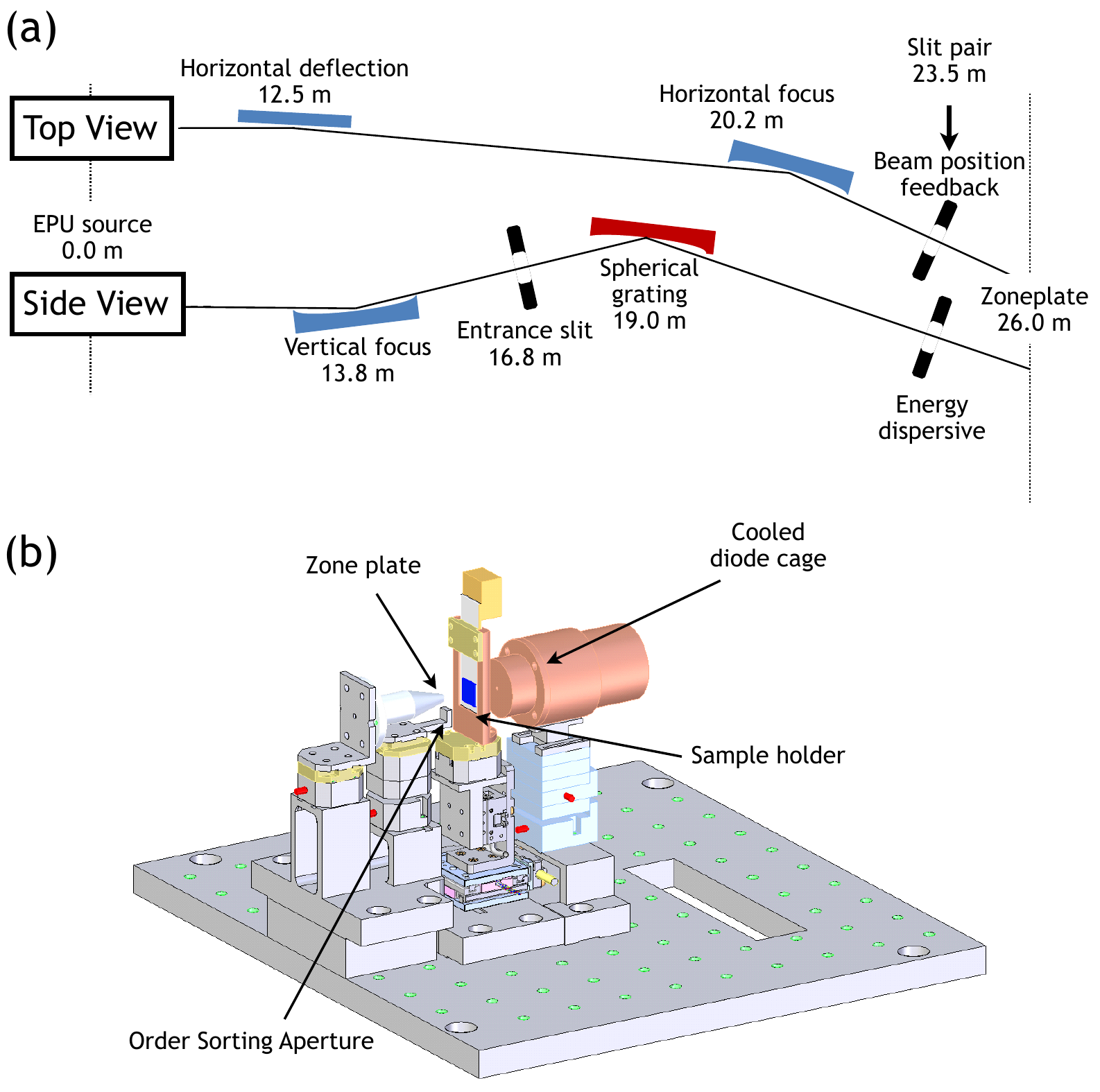}
\caption{(a) Schematic of beamline 13 at SSRL, showing the x-ray optical components from a top-view (top) and side-view (bottom) perspectives. (b) Mechanical drawings of the scanning transmission x-ray microscope.}
\label{fig:Setup}
\end{figure}

The scanning transmission x-ray microscope uses the elliptically polarizing undulator (EPU) at beamline 13 at SSRL as an x-ray source. High energy electrons are stored in a closed orbit at the synchrotron. For this purpose the total stored charge is split and compressed in so-called {\em buckets}. The particular layout of SSRL is compatible with 372 buckets filled with about 1 pC of charge at a kinetic energy of 3 GeV. Each electron bucket is 50 ps long (FWHM) and subsequent buckets are separated by 2.1 ns (476.2 MHz). As an electron bucket traverses the EPU it is forced by variable and tunable magnetic fields on an undulating motion either in the vertical, horizontal plane or both of these, which causes the generation of linear or circular polarized x-rays \cite{carr1993adjustable, JSY:JSYCH2005}.The beamline 13 EPU \cite{carr1993adjustable} produces polarized x-rays in the energy range between approximately 200 eV to 1200 eV. This energy range -- commonly referred to as soft x-ray region -- has been shown to be very well suited to study magnetic properties of complex magnetic nanostructures based on $3d$ and $4f$ transition metals \cite{stohr2007magnetism}.

The diverging x-ray beam generated in the EPU is guided to the microscope using appropriate x-ray mirrors as shown schematically in Figure~\ref{fig:Setup}(a). Note that the typical angle of incidence for soft x-ray optics is a few degree. To obtain monochromatic x-rays at the sample a spherical grating monochromator (SGM) is used in this case. First the x-ray beam is focused on the entrance slit, about 15 micrometer in size, using a spherical mirror. The spherical grating produces a 2:1 magnified image of the entrance slit onto the exit slit, which is located in the energy dispersive plane so that rocking of the SGM changes the energy passing through the exit slit. The typical energy resolution is $E/\Delta E=7000$. In the horizontal plane the beam is deflected first at a flat horizontal mirror that also serves as a power filter, removing higher energy x-rays from the downstream spectrum and thus reducing thermal load on the remaining optical components. Since the source is shared between different experiments at beamline 13 another horizontal mirror is used to deflect the beam towards the experiment. This mirror produces a de-magnified image of the source onto the horizontal exit slit, typically about 50 to 100 micrometer in size. Each slit of the horizontal exit slit assembly is electrically isolated and the difference in electron yield current is used to feedback onto the angle of the horizontally focusing mirror, so that the horizontal beam position can be kept constant independent of beam movements of the source. Once the beam passes through the exit slit, which is the common focus in the horizontal and vertical plane, it diverges again and finally illuminates the zoneplate used to focus the x-rays in the STXM instrument at 26 meters from the source. The beam size at this point is 2.0 mm x 1.0 mm, while the zoneplate (XRADIA, 160 $\mu$m diamter, 80 $\mu$m inner stop, 30 nm outermost zonewidth) is much smaller. Due to the significant overfilling of the zone plate, the setup becomes less sensitive to small remaining beam motions, while still preserving an acceptable photon flux onto the sample. The transmisison of the zone plate is optimized for a photon energy of 800 eV, which is close to the absorption resonances of the ferromagnetic $3d$ transition metals Fe (706.8 eV), Co (778.1 eV) and Ni (852.7 eV). Using a calibrated x-ray photodiode (AXUV-100G, Opto Diode), we measured a photon flux through the zone plate of $2\times10^{10}$ photons/s at 800 eV under normal operating conditions.

The STXM instrument consists of a zone plate as described above and an order sorting aperture (OSA). As OSA we used a standard transmission electron microscopy PtIr(95:5) strip aperture (Hitachi 67137), with 30, 50, 70 and 200 $\mu$m apertures, while for most experiments the 50 $\mu$m aperture is used. STXM instruments are common nowadays at synchrotron source and commercially available. Rather than modifying a commercial microscope to accommodate the special requirements necessary for studies of magnetization dynamics, we developed a dedicated instrument at SSRL. Commercial microscopes rely on interferometer control of the sample position \cite{kilcoyne2003interferometer} to achieve sub 10 nm resolution and stability. However, the use of an interferometer in an UHV environment and in combination with an electromagnet in close proximity to the sample is extremely challenging. For this reason we decided to achieve the a sufficiently high stability by minimizing the size of the setup and keeping the mechanical path lengths between the different optical components short. The complete setup as mounted on an in-vacuum breadboard is shown in Figure~\ref{fig:Setup}(b).

The total length and height of the setup is about 10 cm each and the width is 2.5 cm. The zone plate is moved along the beam direction using piezo stepper stages with resistive encoders (ATTOCUBE ANPx101/RES), and the order sorting aperture can be moved in the plane perpendicular to the beam using similar stages (Attocube ANPx101/RES and ANPz101/RES). Each of these stages allows for 5 mm of motion, which is sufficent for alignment. The detector is mounted in a copper shield that provides electric shielding and temperature stabilization for the avalanche photodiode. The detector is also mounted on piezo stepper motors like the OSA and ZP. The samples are typically mounted on small boards (15 mm $\times$ 40 mm)with incorporated wave guides and SMA connectors. Since the SMA cabling capable of transmitting electric RF signals up to 18 GHz is relatively stiff we decided to use robust piezo motor actuated roller bearing stages (PPX-20 and PPX-32, MICOS) to control the sample position. These also provide a larger range of motion (18 mm horizontal and 12 mm vertical), which is convenient for alignment purposes and allows to potentially mount more than one sample at once.

The sample can be placed into different magnetic fields, using either permanent magnets or electromagnets. Permanent magnets are used to apply up to 0.8~Tesla perpendicular to the sample surface (assuming that the x-rays are incident normal to the sample). A water cooled electromagnet with a gap of 9.0 mm is available to apply magnetic fields in plane up to 0.25~Tesla. The sample can be rotated around its vertical axis to change the projection between incoming x-rays and magnetization and allow the experimenter to detect different components of the magnetization. The entire microscope, which is housed in a stainless steel vacuum chamber, can be aligned with respect to the x-ray beam using six different struts (not shown). All microscope components used are vacuum compatible and using an ion pump we are able to achieve a base pressure of $2\times10^{-8}$ mbar during imaging experiments. Operating the microscope in vacuum is advantageous since it avoids contamination of the surface with hydrocarbons during imaging which is important because many of the experiments require long exposure and averaging time. Even after several days of exposure of the same sample we have never observed a decrease in transmission of more than 50\%, typically obtained in a few seconds or minutes if operated in atmosphere. 

\begin{figure*}[t!]
\centering
\includegraphics[width=\textwidth]{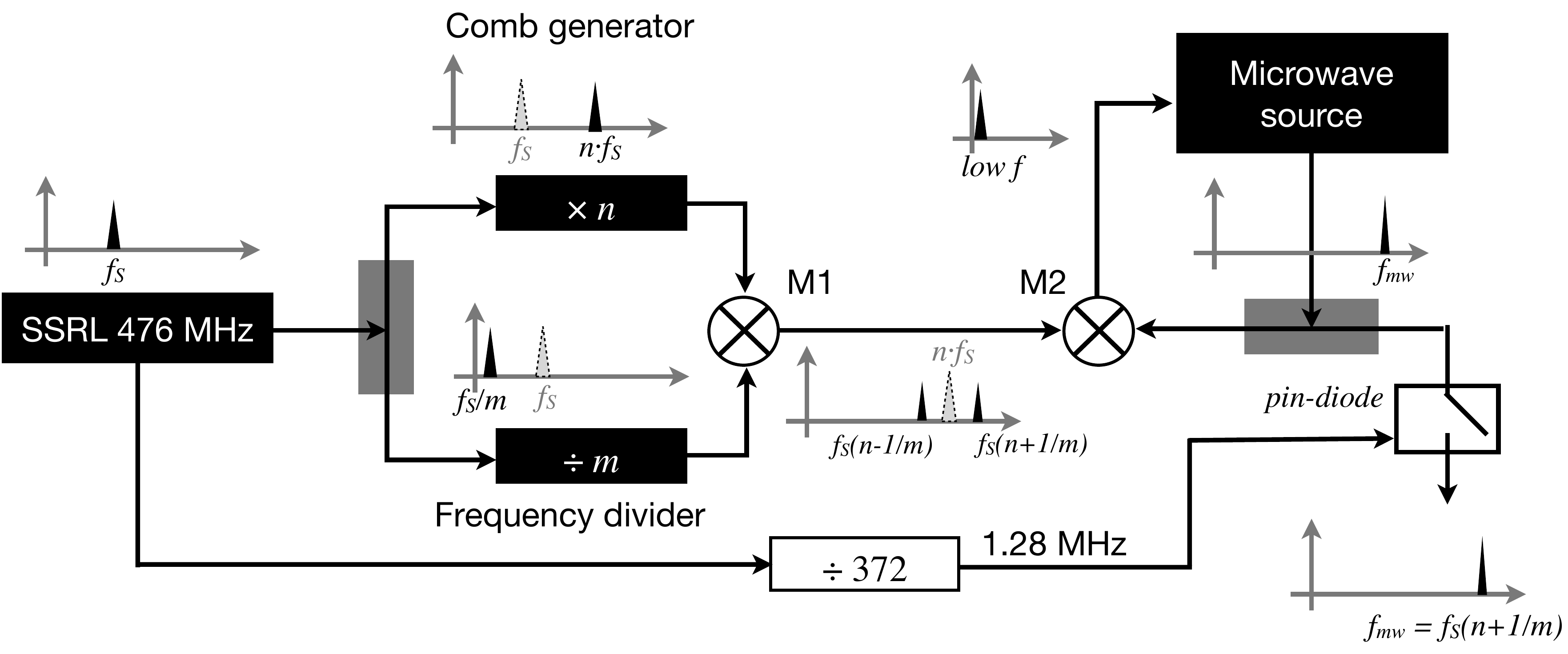}
\caption{Schematic of the microwave synchronization board, described in detail in the text.}
\label{fig:pump:scheme}
\end{figure*}

\begin{figure}[b]
\centering
\includegraphics[width=0.5\textwidth]{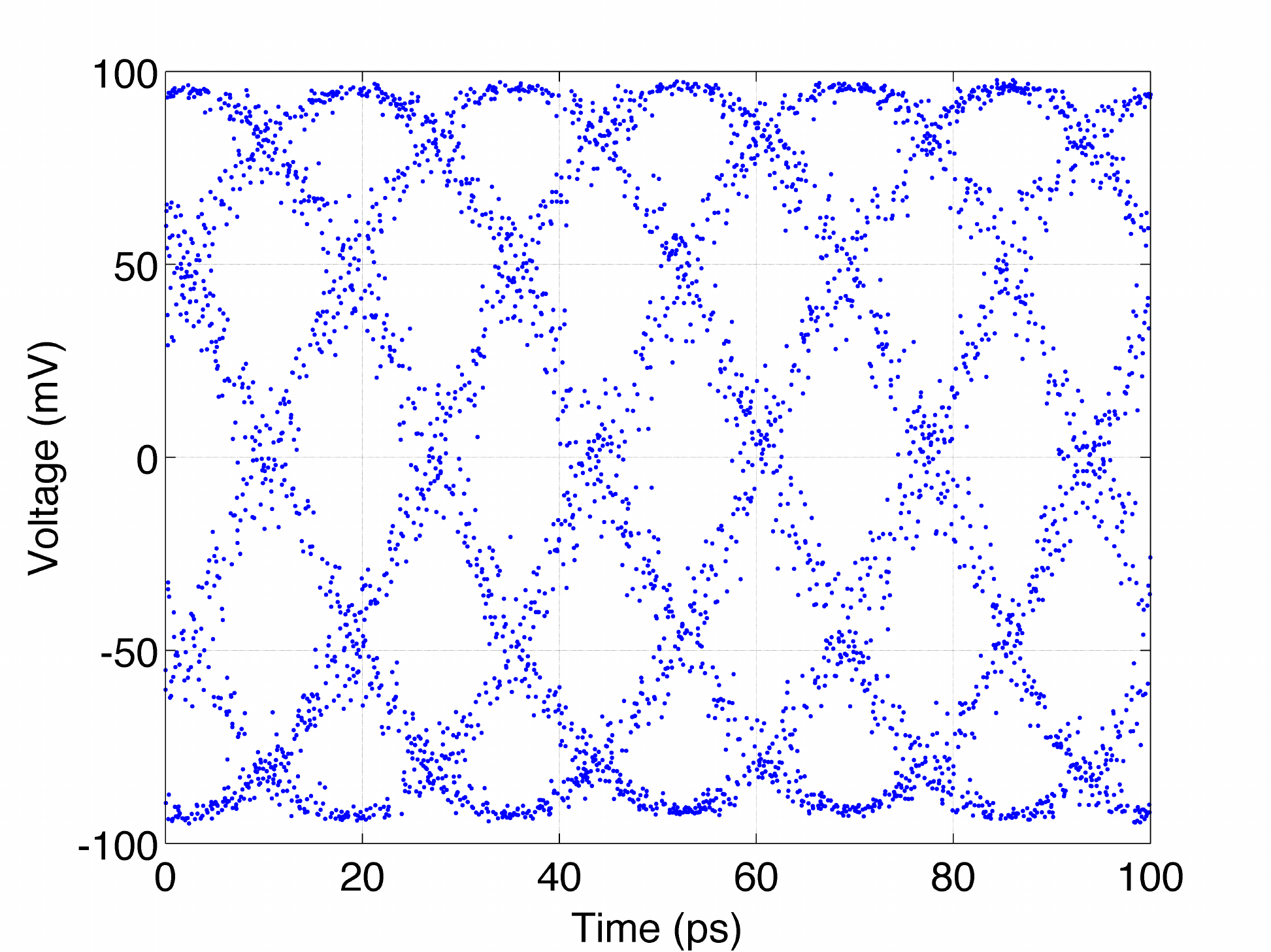}
\caption{Sampling oscilloscope single acquisition run when the synchrotron frequency is used to trigger the measurements of the output of the microwave synhcronization board at $f_{mw}=f_s(20+1/6)=10.07$ GHz.}
\label{fig:sampling_scope}
\end{figure}

\subsection{Single-photon counting scanning transmission x-ray microscopy}

The ``probe'' part of the instrument is a detection system able to look at the individual x-ray pulses generated by all the electron buckets present in the SSRL storage ring. The operating frequency of the SSRL storage ring $f_s\approx476$ MHz, meaning that the electron buckets (and in turn the x-ray pulses) are temporally separated by 2.1 ns. To detect x-ray pulses so closely spaced in time, a special photon-counting electronics was only recently developed\cite{RSI.78.014702}. Such electronics looks at the signal coming from a fast avalanche photodiode (Hamamatsu S12426 Si-APDs) and performs three operations.

Firstly, it digitizes the APD signal using a discriminator synchronized with the synchrotron frequency $f_s$. This is necessary because the number of x-ray photons reaching the APD is low: at the highest intensity through a typical sample, we estimate that on average 1 every 4 electron buckets results in one photon being transmitted through the sample.

Secondly, it sorts all the photons detected by the discriminator into up to 16 different counters, which can be then accessed independently. This is achieved using field-programmable gate-array (FPGA), as described in details in Ref.~\cite{RSI.78.014702}.

Finally, the electronics can be synchronized with the period of the orbit of the storage ring $f_s/N_{buckets}=1.28$ MHz, since $N_{buckets}=372$ at SSRL. Using this feature, half of the counters can be used to measure the x-ray signal during the odd orbits, and the other half during the even orbits. This allows for designing a very effective normalization scheme based on a modulation technique, where the ``pump'' is fed to the sample only during odd revolutions of the ring, and even revolutions are used as the reference signal. Such capability is particularly useful in a synchrotron light source such as SSRL, where the current of the storage ring is kept constant via top-up injections of electrons. Each injection lasts for about 10 seconds and is repeated every 5 minutes. The injections appreciably offset the charge of some of the buckets, which in turn results in an offset of the x-ray intensity of the order of 1\%. This has the potential of completely masking real signals which, as we will show below, can be variation from the background as small as 0.1\%. However, by collecting the signal and the reference from an electron bucket that is almost identical after one revolution period (it takes $1/$1.28 MHz$^{-1}\approx780$ ns for one electron buckets to orbit around the ring) these offsets normalize out in an efficient way, and allows us to reach a noise level of $10^{-4}$, enough to clearly detect the signal.

\subsection{Microwave generation phase-locked to the synchrotron light source}

The goal of the microwave synchronization board is to generate a clean microwave signal at a single frequency phase locked to the synchrotron frequency. The microwave signal can be used to drive excitations in the sample, and these excitations will be in turn phase locked to the synchrotron frequency.

The peculiarity of our microwave synchronization board is that it is phase-locked not to an exact harmonics $n$ of the synchrotron frequency, but to a frequency that is offset from it by a subharmonics $f_S/m$, where $m$ is an integer. This is an approach similar to the one followed in Ref.~\cite{PhysRevLett.106.167202}. When this situation is realized, the measurement is not longer a pure stroboscopic type of detection, because two consecutive x-ray photons will not measure the same phase of the excitation. However, the phase observed by the photons is not random: every $m$th photon will always see the same phase. Our measurement is ``quasi''-stroboscopic in the sense that a measurement does not record only one specific phase of the excitation, but a finite subset of phases in each pixel before moving to the successive one. By collecting all the phases almost simultaneously, we completely suppress even the slight long term drift, which is inevitably left in our scanning stages without interferometric control.

Fig.~\ref{fig:pump:scheme} shows the schematic of the microwave synchronization scheme. The synchrotron clock frequency $f_s\approx476$ MHz is split with a 3 dB power divider into two parallel transmission lines. Along the first one, a frequency comb generator is inserted in order to obtain simultaneous generation of $n$ harmonics of $f_s$, with $2\leq n \leq 24$. In the second transmission line, a frequency divider is mounted in order to produce a single $f_s/m$, where $m$ can be programmed to have a value between 2 and 17.

The two transmission lines carrying the $n$ harmonics and, respectively, one subharmonics of $f_s$ are recombined using the frequency mixer M1 to produce modulation sidebands located at the desired $f_{pump} = f_s(n\pm1/m)$. This signal is already in phase with $f_s$, but it also contains 23 other harmonics of $f_s$, some of which may act as a further pump frequency for the sample and disturb the measurement. In order to excite the sample with a clean signal at the single frequency $f_{pump}$, we introduce a low phase-noise microwave signal generator (Anritsu MG3692B) into the circuit.

The frequency $f_{mw}$ of the microwave signal generator can be tuned to be close to $f_{pump}$ with a beating of a few Hz. 
The low frequency beating signal is now a measure of the phase difference between the two signals $f_{pump}$ and $f_{mw}$. Using a PID controller (Stanford Research Systems SIM960) and connecting its output to the electronic frequency control (EFC) port of the microwave signal generator\footnote{This port allows steering of the internal clock of the signal generator by an amount $\Delta f = \alpha V_{PID}$, where $V_{PID}$ is the output of the PID controller and $\alpha=8\cdot10^{-8}$ Hz/V} we then realized a phase locked loop (PLL), that minimizes the phase difference between the two signals. When the PLL is closed, $f_{mw} = f_{pump}$, and a single frequency signal locked to the synchrotron frequency $f_s$ is provided to the sample.

The successful implementation of the synchronization scheme can be checked with a sampling oscilloscope (Tektronix TDS8200 with 80E04 sampling module) where the trigger port is connected to the synchrotron reference signal at $f_s=476$ MHz and the sampling port to the output of the microwave synchronization board. For this measurement, we used $n=21$ and $m=6$, to get $f_{mw}=10.07$ GHz. We chose this specific frequency to prove that our microwave synchronization board works reliably up to the higher end of 5-10 GHz frequency range. Taking a series of single acquisition shots of the scope trace produces consecutive output traces similar to the one plotted in Fig.~\ref{fig:sampling_scope}. The six phases of the signal at a frequency very close to 10 GHz are clearly distinguishable on the screen of the sampling scope, and the fact that consecutive single shots do not drift on the screen proves that there exists a well-defined phase relation between the synchrotron frequency and the output of our board.

In the actual experiment, the microwave signal is fed to the sample only during the odd orbits of the storage ring, and turned off during the even orbits, as mentioned in section IIA. This is achieved using a fast pin-diode, a device able to switch on and off the transmission of a microwave signal within $\sim$10 ns, much faster than the storage ring orbit period of 780 ns.

\begin{table*}[t!]
\begin{tabular}{|l|c|c|c|c|c|c|c|c|c|c|c|c|c|c|}
\hline
& & & & & & & & & & & & & & \\
$n$ (harmonics) & 11 & 12 & 13 & 14 & 15 & 16 & 17 & 18 & 19 & 20 & 21 & 22 & 23 & 24\\
& & & & & & & & & & & & & & \\
\hline
& & & & & & & & & & & & & & \\
$nf_s + f_s/6$ (GHz) & 5.32 & 5.79  & 6.27 & 6.74 & 7.22 & 7.70 & 8.17 & 8.65 & 9.12 & 9.60 & 10.07 & 10.55 & 11.03 & 11.50\\
& & & & & & & & & & & & & & \\
\hline
& & & & & & & & & & & & & & \\
$J$ (fs) & 280 & 260 & 290 & 280 & 320 & 320 & 310 & 340 & 370 & 430 & 480 & 960 & 1120 & 1340\\
& & & & & & & & & & & & & & \\
\hline
\end{tabular}
\caption{Tabulated values of the measured jitter $J$ for different harmonics $n$ of the comb generator at $m=6$. The generated frequency $f_{mw}$ as well as it corresponding period $T$ are also indicated.}
\end{table*}

\begin{figure}[t]
\centering
\includegraphics[width=0.5\textwidth]{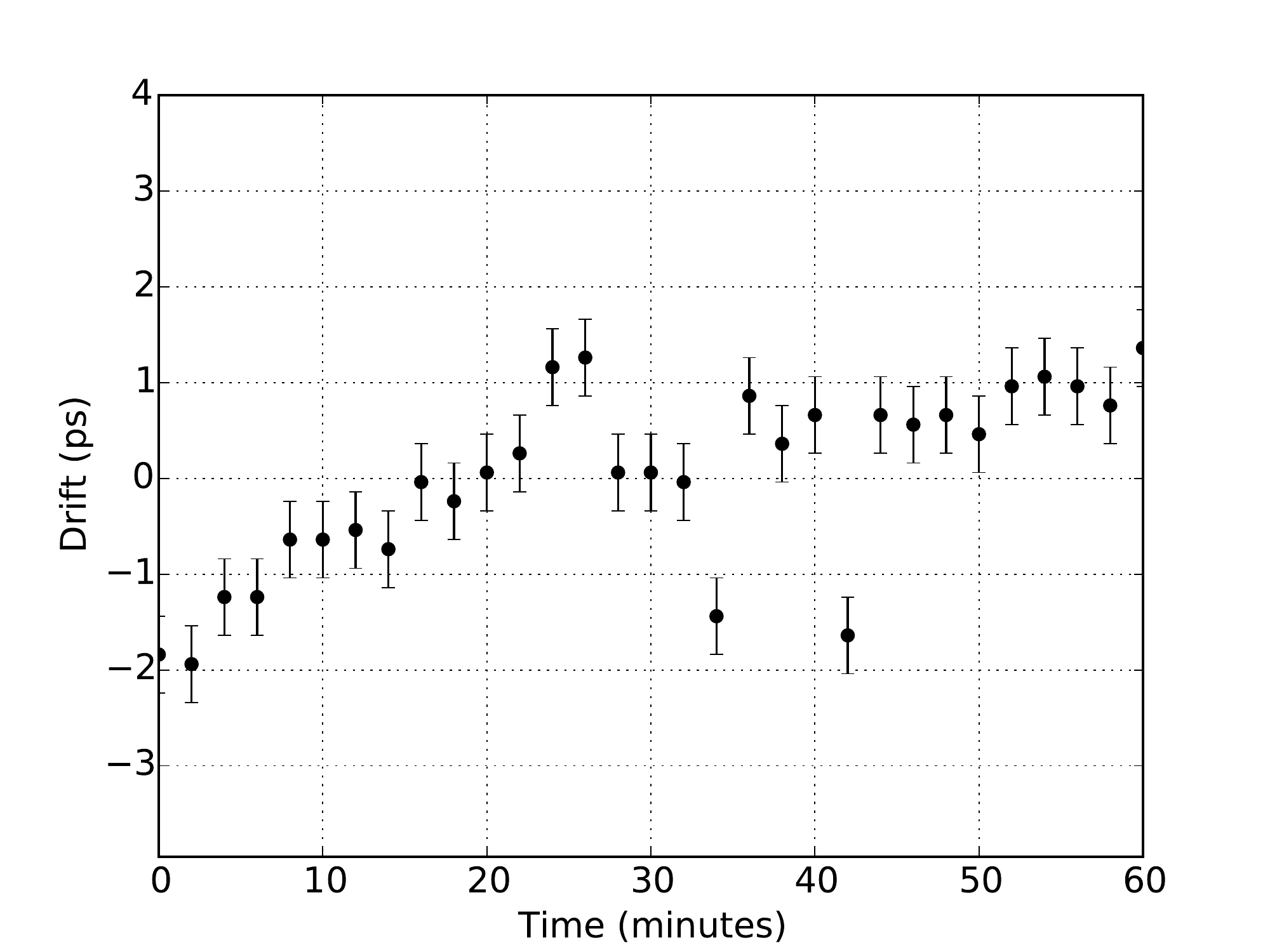}
\caption{Maximum drift of the microwave synchronization board over a measurement period of one hour for at a frequency $f=7.69$ GHz ($n=16$, $m=7$).}
\label{fig:drift}
\end{figure}

\section{Characterization of the setup}

In this section, we quantify the robustness of the synchronization scheme, both in terms of jitter and drift.

In order to quantify the in-loop jitter $J$, we use the output of mixer M2, which contains the beating between the signal from the microwave source and the one from the comb generator. This beating is a direct measure of the phase difference between the two signals, and one only needs to normalize it in the proper units to retrieve the in-loop timing jitter. This is done by measuring the amplitude of the beating for the cases of open and closed PLL. Then, one can use the relation
\begin{equation}
J = \dfrac{V_{closed}/V_{open}}{2\pi f},
\label{eq:jitter}
\end{equation}
where $V_{closed}$ is the rms voltage measured for the closed loop, $V_{open}$ measures the amplitude of the beating oscillation in open loop. The expression is valid for small phase fluctuations compared to the oscillation period, i.e. $\sin\phi(t)\approx\delta\phi(t)=J\, 2\pi f$. In Table I we list the calculated jitter for the different harmonics $n$, and for a single value $m=6$ of the subharmonics. It is clear that for all harmonics the jitter is of the order of 1 ps, and below 500 fs for frequencies up to 10 GHz.

We now turn to the estimation of the drift of the microwave synchronization board. In order to do this, we used a second board consisting of three parts: a frequency multiplier that generates the 16th harmonics of $f_s$ at 7.62 GHz, a frequency divider that produces $f_s/7=68$ MHz, and a mixer that can be used to produce a modulation signal at 7.69 GHz. This modulation signal can then mixed down with the nominally identical signal generated by our microwave synchronization board programmed with $n=16$ and $m=7$. Similarly to the jitter measurements, the down-mixing voltage is a measure of the phase difference between the two signals, and can be used to estimate the drift of the board.

This method actually only allows to estimate the upper limit of the drift, because the contribution from the two boards cannot be disentangled. For simplicity, we assumed that our board was responsible for all the drift observed in the measurement. We also note that we did not implement any temperature stabilization, since this may be unpractical in a real experiment. Fig.~\ref{fig:drift} plots the drift over 1 hour long measurement, showing the timing drift is bounded between $\pm2$ ps over the hour-long time span.

Both timing jitter and drift values are much smaller than the temporal FWHM of the x-ray photons, which in the low-emittance operation mode is about 50 ps. Hence, we conclude that our microwave synchronization board board does not cause any significant degradation of the temporal resolution available at the synchrotron.

%

\begin{figure}[b]
\centering
\includegraphics[width=0.5\textwidth]{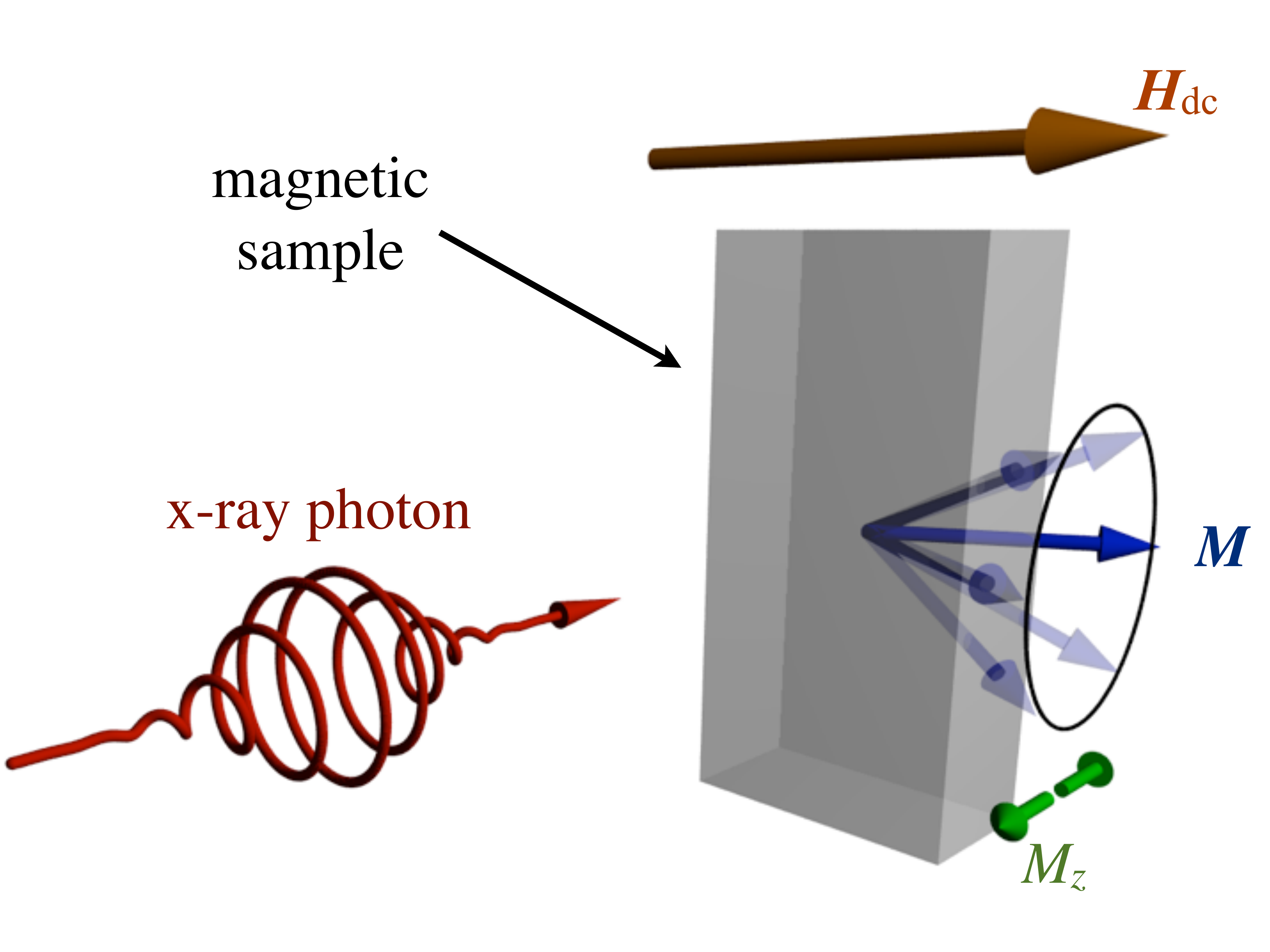}
\caption{Schematic of the phase-resolved measurement of the magnetization dynamics. Circularly polarized x-ray photons tuned at the $L_3$ resonant edge of Ni are used to measure the phase dependent XMCD contrast arising from the precessing magnetization in the sample.}
\label{fig:fmr_schematic}
\end{figure}

\begin{figure*}[t!]
\centering
\includegraphics[width=\textwidth]{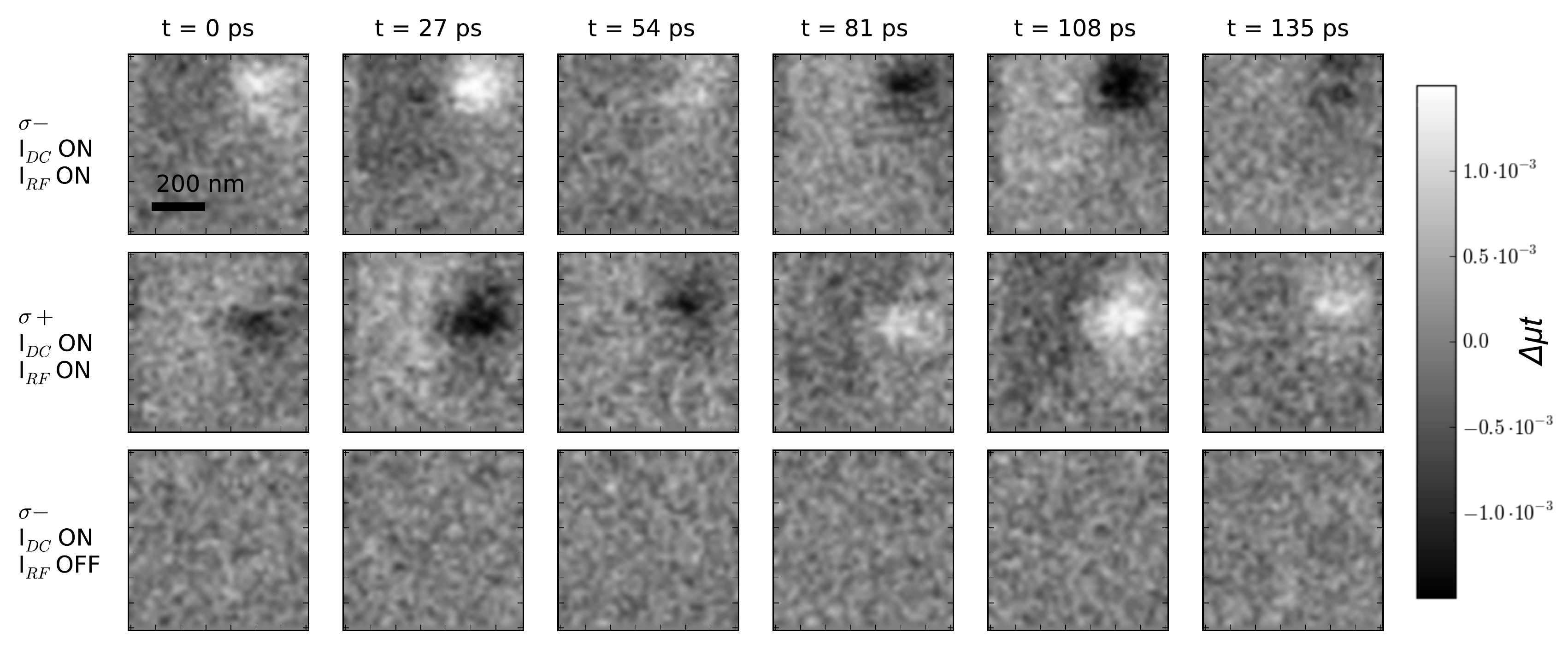}
\caption{Time-resolved XMCD images of the magnetization dynamics in a spin torque oscillator. Each row shows a set of six images representing six subsequent phases of the magnetization precession. Data measured for negative (first row) and positive (second row) helicity of the x-rays and with both $I_{DC}$ and $I_{mw}$ applied to the sample. (Third row) Data recorded for negative x-ray helicity when no microwave current $I_{mw}$ was applied to the sample.}
\label{fig:fmr_images}
\end{figure*}

\section{Applications}

In order to demonstrate the capabilities and the versatility of our setup to measure at high-frequency and with high-spatial resolution we show the results from two types of samples. At first, we mapped the large amplitude spin waves excited by a spin polarized current in a 5 nm think NiFe film. The frequency of the spin wave was about 6 GHz, and the excitation extended for an area of the order of 100 nm in size. Then, we measured small angle ($\sim0.1$ degrees) magnetization precession due to ferromagnetic resonance in a Co microstrip (20 nm thick), driven by a microwave magnetic field at a frequency of about 9 GHz.

For both cases, the geometry of the experiment is shown in Fig.~\ref{fig:fmr_schematic} in a schematic way. A static magnetic field $H_{dc}$ is applied along the plane of the film. This defines the precessional axis of the magnetization (parallel to the applied field) so that during the precession there will be a non-zero component of the magnetization parallel to the x-ray direction ($M_z$ in our notation).
This is a necessary condition for the x-ray magnetic circular dichroism effect (XMCD) \cite{stohr2007magnetism} to produce a contrast, depending on whether the spins are parallel or anti-parallel to the helicity of the x-ray photons. The helicity of the x-rays is controlled by the EPU that delivers x-rays to beamline 13-1 at SSRL, as described in Section IIA. The undulator in combination with a monochromator also allows to finely tune the energy of the x-ray to the desired value.

\subsection{Nanoscale spin wave dynamics in spin torque oscillators}

The time-resolved mapping of spin waves excited in a spin torque oscillator has remained so-far an elusive goal for the magnetism community. In such a device, a high-density, spin-polarized direct current is injected via a nanometer-size contact into an extended ferromagnetic film. Via the spin transfer torque effect \cite{berger1996emission,slonczewski1996current}, spin waves are excited in the region surrounding the nanocontact. The spin waves can also be efficiently phase-locked to an external microwave signal superimposed to the driving direct current \cite{PhysRevLett.95.067203}. Both these aspects, the nanoscale characteristic of the excitation and the possibility of phase-locking with an external signal, make this sample the ideal testing ground of our setup. The details of the sample fabrication can be found in Ref.~\cite{demidov2010direct} and in our upcoming publication\cite{bonetti2015arXiv}. Here we only note that the sample is grown on a SiN membrane transparent to x-rays. Such substrate allows for detection of the x-rays transmitted through the sample via an avalanche photodiode, as required by our detection scheme. Since the spin waves are excited in a thin permalloy (Ni$_{80}$Fe$_{20}$) film, we tuned the x-ray energy to the Ni $L_3$ absorption edge ($E = 852.7$ eV) to maximize the XMCD contrast.

The spin wave dynamics is first excited via direct current provided to the sample with a precision current source (Keithley Source Meter 2400), and the frequency of the excitation observed in real-time using a spectrum analyzer (Rhode\&Schwarz FSU13). By varying the current magnitude, one can accurately control the frequency of the spin wave and bring it close to one of the frequencies that can be generated by our microwave synchronization board. For the data present below, we found that a direct current $I_{DC}=8$ mA produces spin waves at a frequency of 6.27 GHz, very close to the $n=13$, $m=6$ line of the microwave board. \footnote{The choice of $m=6$ rather than $m=8$ (the maximum number of counters of our detectors) was dictated by the requirement that the number of buckets in the storage ring $N=372=2\cdot2\cdot3\cdot31$ should be divisible by $m$.}

The signal from the microwave board is then fed to the sample superimposed to the direct current to induce synchronization between the spin wave excitation and the synchrotron. When the synchronization condition is satisfied, successive circularly polarized x-ray photons will probe six subsequent projection $M_z$ of the magnetization vector {\bf M} along the x-ray propagation direction. This realizes a phase-resolved detection of the magnetization dynamics which can be used to build a real-space, dynamical map of the spin waves excited at the nanoscale, i.e. a spin wave ``movie''.

The results are presented in Fig.~\ref{fig:fmr_images}. In each row, the images corresponding to the six different phases are plotted in sequence, and the corresponding time delay relative to the first image is indicated above each image in the first row. The contrast arises because of varying x-ray transmission through the sample, with the brighter (darker) colors indicating regions of higher (lower) transmission. Each image is 0.6 $\mu$m $\times$ 0.6 $\mu$m in size and the step size for the scanning sample stage was 30 nm, close to the resolution of 35 nm provided by our zone plates.

The first row shows the data recorded with x-rays having positive helicity and with both direct current ($I_{DC}$) and microwave signal ($I_{mw}$) applied to the sample. On the top-right corner, a white contrast is observed in the first image at $t=0$ ps. The contrast becomes larger in the second image at $t=27$ ps, it weakens in the third image and it turns black in the fourth. The black contrast undergoes a similar evolution in the fourth, fifth and sixth image ($t=81$, $108$ and $135$ ps respectively) as the white contrast in the first three images.
The contrast is due to magnetic dichroism, hence the opposite contrast has the meaning of the magnetization in the sample pointing in two opposite directions. The topographic features of the sample, such as the nanocontact where the current is injected, are normalized out by the even vs odd orbit normalization scheme described in section IIA. The maximum amplitude of the signal correspond to a variation with respect to the background signal of the order of $10^{-3}$, with the noise fluctuations on the order of $10^{-4}$ in each of the recorded phase.

In order to prove that we are observing a magnetic excitation, we reverse the helicity of the x-rays. In the second row of Fig.~\ref{fig:fmr_images} we plot the images recorded when the same $I_{DC}$ and  $I_{mw}$ were applied to the sample, but now with the x-ray helicity being reversed. It is evident that the contrast has changed sign, proving that the signal is magnetic. The shift of the contrast towards the middle of the image compared to the first row is due to drifting of the microscope stages in the time between the two consecutive measurements, of the order of a couple of hours. This observation further motivates our measurement scheme, insensitive to such drift.

Finally, in the last row we present the images taken when no microwave signal was applied to the sample. The spin wave was still observed in the spectrum analyzer, but it was not phase locked to the synchrotron. In absence of phase-locking, the average out-of-plane component of the magnetization is zero, and uniform gray images are expected and observed. Similarly uniform images are recorded if only the microwave signal is applied to the sample, without any $I_{DC}$ sustaining the spin wave excitation \cite{bonetti2015arXiv}.

\begin{figure}[t]
\centering
\includegraphics[width=0.48\textwidth]{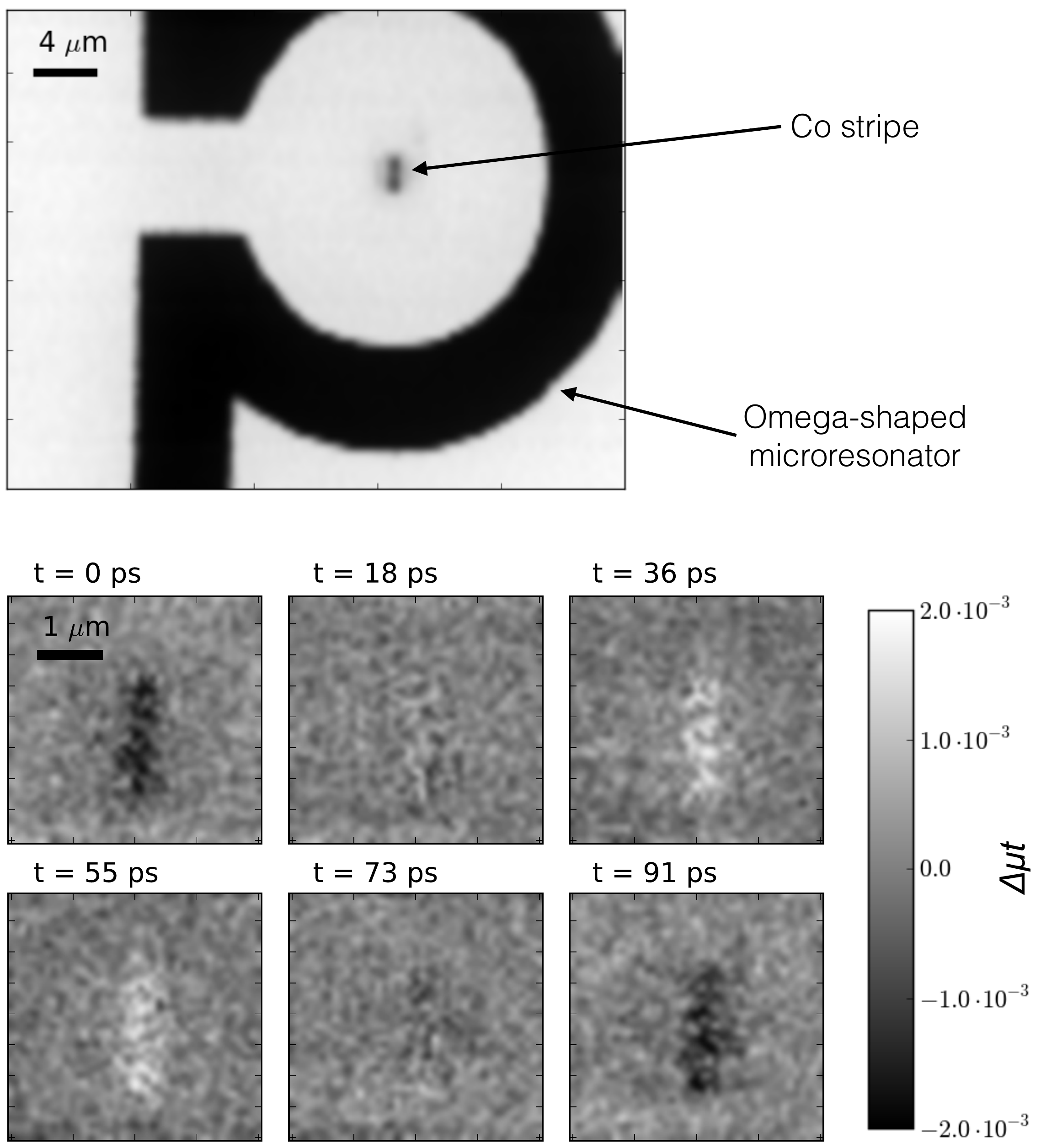}
\caption{(Top panel) Overview of the sample for the x-ray ferromagnetic resonance (X-FMR) measurements. (Bottom panel) Images of a cobalt microstrip excited by an microwave magnetic field at $f=9.129$ GHz applied orthogonal to the image plane. A static magnetic field $\mu_0H=135$ mT was applied parallel to the short side of the microstrip.}
\label{fig:XFMR]}
\end{figure}

\subsection{Ferromagnetic resonance in confined structures}

Ferromagnetic resonance (FMR) in multilayers composed of different materials can result in complex dynamics that cannot be easily untwinned only by measuring magnetic fields. X-ray FMR (X-FMR) solves the problem by exploiting the elemental sensitivity of the XMCD effect. In this specific measurement, we aimed at addressing the upper frequency limit of our synchronization scheme, in combination with 50 ps FWHM x-ray pulses. The sample is a 20 nm thick Co microstrip (2 $\mu$m x 0.5 $\mu$m), that is deposited in the center of an omega-shaped microresonator of several tens of micrometer diameter using optical and electron beam lithography tools\cite{banholzer2011visualization}, as shown in the top panel of Fig.~\ref{fig:XFMR]}. The lithography was performed also in this case on a SiN membrane that allows for x-rays to be transmitted through it. 

An external magnetic field of 135 mT was applied perpendicular to the easy axis of the sample (horizontal, parallel to the short side of the strip in Fig.~\ref{fig:XFMR]}), while the sample was exposed to an RF field at 9.129 GHz. The RF field frequency and applied magnetic field were optimized after conventional ferromagnetic resonance measurements acquired ex-situ. In analogy with the spin wave measurements, the STXM images in Fig.~\ref{fig:XFMR]} show the XMCD contrast at the Co L$_3$ edge ($E=778.1$ eV) arising because of the time-varying magnetization component perpendicular to the plane of the image, as depicted schematically in Fig.~\ref{fig:fmr_schematic}. Between two consecutive images there exists a phase difference of 60 degrees, corresponding to a time difference of 18 ps. Images acquired 180 degree apart show inverted contrast as expected. The fact that we acquire two images, one with microwave on and one without microwave excitation, allows us to directly measure only the change in the absorption cross section caused by the magnetization dynamics, while removing the background fluctuations common to both images.

The data show that the cross section $\Delta\mu t$ changes by $\pm 2\times10^{-3}$. Using reference values for the XMCD we can then estimate the out-of-plane precession angle of the magnetization to be about 0.1 degree, without considering the reduction of the observed contrast due to the mismatch of x-ray pulse length of investigated time scale. The convolution of a 50 ps FWHM gaussian with a sinusoidal curve at 9 GHz frequency results in a sinusoidal curve with approximately 50\% reduced amplitude. Hence, we can conclude that in the studied sample we are able to detect a precession angle of the order of 0.1 degrees.

\section{Conclusions}
We have built a microwave synchronization electronics that combined with a synchrotron based scanning x-ray microscope allows for nanoscale and time-resolved measurements at frequencies up to 10 GHz, beyond the current state-of-the-art instruments. The jitter ($\approx500$ fs) and the drift ($\approx2$ ps over 1 hour measurement) of the board are much smaller than the temporal FWHM of the x-ray pulses ($\approx50$ ps), which is hence the only factor determining the temporal resolution of the setup. We have used a quasi-stroboscopic synchronization scheme to record multiple phases of the magnetization precession in parallel. This has allowed us to virtually eliminate the drift which is otherwise intrinsically present in a scanning microscope.

The illuminating x-ray optics and the in-house designed x-ray microscope are optimized for long term stability and minimal contamination due to its operation in ultra high vacuum. The in-house design also allows for flexible sample environment, which allows for including magnetic fields and electric excitations of various kinds, critical for state of the art experiments addressing dynamical properties at the nanoscale.

A single photon detection system combined with our microwave synchronization board allowed us to record micrometer-sized images with 35 nm resolution within an acquisition time of a few hours. In this time frame, we could reliably measure dynamical signals which were 0.1\% variations over the background. When six different phases were recorded, the typical noise level was of the order of 0.01\%. This has allowed us to create a nanoscale dynamical map of the spin waves excited in a nanocontact spin torque oscillator, as well as of the ferromagnetic resonance in a patterned Co strip.

Our electronic design can be readily extended to frequencies up to 20 GHz, and theoretically to the $\sim$100 GHz range given the availability of microwave components compatible with those frequencies. This is of potential interest in storage rings operating in low-alpha mode \cite{huang2007low}, which can produce 1-10 ps long x-ray pulses, as well as in free electron lasers, where sub-ps pulses are available.

\section{Acknowledgments}
We are very grateful to Sergei Urazhdin at Emory University for fabricating the sample for the spin wave measurements. Research at SLAC was supported through the Stanford Institute for Materials and Energy Sciences (SIMES) which like the SSRL user facility is funded by the Office of Basic Energy Sciences of the U.S. Department of Energy. Stefano Bonetti gratefully acknowledges support from the Knut and Alice Wallenberg Foundation.\\

%

\end{document}